\documentclass[AMA,STIX1COL,courier]{WileyNJD-v2}
 
\newcommand{\refmath}[1]{(\ref{#1})}
\newcommand{\MatrFormat}[1]{{\it #1}}
\newcommand{\MatrFormatGreek}[1]{{\it #1}}
\newcommand{\VecFormat}[1]{{\bf #1}}
\newcommand{\VecFormatGreek}[1]{\boldsymbol{#1}}

\allowdisplaybreaks

\articletype{Research article}%

\begin{document}

\title{Sample size calculation and blinded recalculation for analysis of covariance models with multiple random covariates}

\author[1,2,3]{Georg Zimmermann*}

\author[4]{Meinhard Kieser}

\author[1,5]{Arne C. Bathke}

\authormark{ZIMMERMANN \textsc{et al}}

\address[1]{\orgdiv{Department of Mathematics}, \orgname{Paris Lodron University}, \orgaddress{\state{Salzburg}, \country{Austria}}}

\address[2]{\orgdiv{Department of Neurology}, \orgname{Christian Doppler Clinic and Centre for Cognitive Neuroscience}, \orgaddress{\state{Salzburg}, \country{Austria}}}

\address[3]{\orgdiv{Spinal Cord Injury and Tissue Regeneration Centre Salzburg}, \orgname{Paracelsus Medical University}, \orgaddress{\state{Salzburg}, \country{Austria}}}

\address[4]{\orgdiv{Institute of Medical Biometry and Informatics}, \orgname{University of Heidelberg}, \orgaddress{\state{Heidelberg}, \country{Germany}}}

\address[5]{\orgdiv{Department of Statistics}, \orgname{University of Kentucky}, \orgaddress{\state{Lexington, KY}, \country{USA}}}

\corres{*Georg Zimmermann, Spinal Cord Injury and Tissue Regeneration Centre Salzburg, Paracelsus Medical University, Strubergasse 21, A-5020 Salzburg. \email{georg.zimmermann@pmu.ac.at}}

\abstract[Summary]{
When testing for superiority in a parallel-group setting with a continuous outcome, adjusting for covariates (e.g., baseline measurements) is usually recommended, in order to reduce bias and increase power. For this purpose, the analysis of covariance (ANCOVA) is frequently used, and recently, several exact and approximate sample size calculation procedures have been proposed. However, in case of multiple covariates, the planning might pose some practical challenges and surprising pitfalls, which have not been recognized so far. Moreover, since a considerable number of parameters have to be specified in advance, the risk of making erroneous initial assumptions, leading to substantially over- or underpowered studies, is increased. Therefore, we propose a method, which allows for re-estimating the sample size at a prespecified time point during the course of the trial. Extensive simulations for a broad range of settings, including unbalanced designs, confirm that the proposed method provides reliable results in many practically relevant situations. An advantage of the reassessment procedure is that it does not require unblinding of the data. In order to facilitate the application of the proposed method, we provide some \texttt{R} code and discuss a real-life data example. }




\maketitle


\section{Introduction}\label{Intro}

In clinical trials with a 
parallel-group design, the main interest is often focused on the question whether a particular intervention is more efficacious than a competitor or placebo. For continuous outcomes, the alternative hypothesis of superiority can be examined by using a one-sided two-sample t test. However, it is well known that in general, adjusting for one or several covariates may increase the power of the test and reduce the bias of the effect estimator.\citep{Hui} For example, it is sensible to include at least one baseline measurement of the outcome as a covariate in the statistical model.\citep{Fri92} The importance of an adjustment for baseline measurements has been highlighted and discussed extensively in a recent regulatory guideline.\citep{Ema15} We would like to emphasize that settings where more than one random covariate should be accounted for, are supposed to be frequently encountered in applied research. For example, in order to assess the efficacy of a treatment in spinal cord injury patients with respect to bladder function \citep{Sug17}, it may be sensible to adjust not only for the baseline measurement of the outcome (e.g., detrusor pressure), but also for another baseline variable, which is presumably correlated with the outcome (e.g., cystometric volume). In a recently published clinical trial from stroke research, the NIHSS score at 24 hours after the intervention was considered as the primary outcome, and group means were adjusted for NIHSS at baseline \citep{Sch16}. However, incorporating, for example, the age of the patient as a further random covariate into the model would have been an appealing alternative way of analyzing the data from that trial. \\

The analysis of covariance (ANCOVA) is a well-established statistical method, which allows for covariate adjustment. 
ANCOVA tests have been used to assess treatment effects in  clinical trials from virtually every medical research area (e.g., neurology \citep{How17,Spe17}, gynaecology \citep{Gho18}, geriatrics \citep{Hai17}). Applying an appropriate method for sample size calculation plays a key role in the planning phase of any interventional study. It has to be ensured that the target power is indeed achieved or, on the other hand, that patients are not unnecessarily exposed to potentially harmful treatments.\citep{IchE9,Moh10} However, a recent systematic review indicates that even if the ANCOVA was used, sample sizes were not calculated appropriately.\citep{Das13} This might reflect the lack of awareness that the topic of sample size calculation in ANCOVA models has recently been addressed in several publications.
Tereenstra et al. considered an ANCOVA model with a single baseline measurement of the outcome in the context of a cluster-randomized trial design.\cite{Ter12} Sample size calculation in the classical parallel two-group setting was discussed in a very concise and readily understandable way by Borm et al.\cite{Bor07} They exploited some analogies between the t-test and the ANCOVA test statistics, in order to get two approximate formulas, which were based on the classical normal approximation and the Guenther-Schouten adjustment, respectively.\citep{Gue81, Sch99} Recently, however, this approach has been criticized by Shieh, who derived an exact method for calculating power and sample sizes for an ANCOVA model with multiple random covariates.\cite{Shi17} Tang proposed another two methods and compared them to existing approaches in the context of designs with and without stratification.\cite{Tan18} Apart from these papers, the seminal publication of Frison and Pocock \citep{Fri92} as well as a recent approach, which is based on sample size calculation methods for multivariate outcomes \citep{Chi18}, should be mentioned as references for the special case of an adjustment for several repeated measurements of the outcome variable.\\
 
With an increasing number of covariates, the amount of uncertainty in sample size calculation increases, too. In the ANCOVA model, additionally to the variance of the outcome and the effect size, one has to provide ``good guesses'' of the covariance matrix of the covariates as well as the correlations between the outcome and the covariates. If these hypothesized values, which can be either based on previous studies or subject matter expertise, are not close to the true values, the resulting actual power and calculated sample sizes might be inaccurate. Therefore, it would be sensible to recalculate the sample size at some pre-specified time point during the course of the study. 
This can be regarded as a special case of an adaptive design, although it should be noted that only the nuisance parameters, but not the effect sizes are re-estimated in the interim analysis.\citep{Bau15,Was} A crucial issue with that sort of recalculation methods is whether or not blinding of the data is required. If unblinding is needed for the interim analysis, a data monitoring committee has to be established, in order to preserve the integrity of the study. Since this is resource-consuming, an appealing alternative would be to use a sample size reassessment method, which does not require unblinding of the data. For an overview of existing methods, we refer to the comprehensive reviews of Proschan\citep{Pro05} and Friede and Kieser.\cite{Fri06}\\

In the present paper, we propose a blinded sample size recalculation procedure for an ANCOVA model with multiple random covariates, extending the methods that have been examined by Friede and Kieser for the case of one single random covariate.\citep{Kie11} Our manuscript is organized as follows. In Section \ref{FormulasRealLife}, we introduce some notation as well as sample size formulas for the fixed design, and describe the main steps of our proposed recalculation method in detail. Moreover, we also place emphasis on practical aspects by discussing some potential problems, which might arise when (re-)calculating sample sizes for a multiple ANCOVA model, and suggest appropriate remedies. To our knowledge, these surprising difficulties have not been addressed in the literature so far. Then, in Section \ref{Simu}, we investigate the performance of both fixed sample size calculation formulas and our proposed recalculation procedure with respect to type I error rates, empirical power and average as well as maximum sample sizes in an extensive simulation study, covering a broad range of parameter configurations. We demonstrate the application of our proposed method to real-life data in Section \ref{RealLife}. We conclude with a discussion of the advantages and limitations as well as some ideas for future research. In the online supplement, we provide the \texttt{R} code that can be used for applying our proposed method in real-life settings, as well as the simulation results for the fixed sample size calculation settings.

\section{Sample size formulas, blinded sample size recalculation procedure, and practical considerations}
\label{FormulasRealLife}

\subsection{Approximate formulas for the fixed sample size setting}

Let $(Y_{ij},{\bf Z_{ij}'})$ be independent, following a multivariate $(c+1)$-dimensional normal distribution with mean vector $(\mu_i, \VecFormatGreek{\mu}_Z')'$ and covariance matrix $\MatrFormatGreek{\Sigma}$, $1\leq j \leq n_i$, $i \in \{1,2\}$. The adjusted means $\mu_1, \mu_2$ may differ, whereas the covariate means and the covariance matrix are assumed to be equal across the groups. The covariance matrix $\MatrFormatGreek{\Sigma}$ of the joint distribution of the outcome and the covariates can be conveniently expressed as
\[
\MatrFormatGreek{\Sigma}:= 
\begin{bmatrix}
\sigma_Y^2 & \VecFormatGreek{\sigma}_{YZ}^{\prime}\\
\VecFormatGreek{\sigma}_{YZ} & \MatrFormatGreek{\Sigma}_Z
\end{bmatrix},
\]
where $\sigma_Y^2$ and $\MatrFormatGreek{\Sigma}_Z$ denote the variance of the outcome and the covariance matrix of the covariates, respectively, and $\VecFormatGreek{\sigma}_{YZ}$ represents a $c$-dimensional vector, which contains the covariances between the outcome and each covariate. \\
Consider testing for superiority of group 1 over group 2 in terms of the group means, that is, $H_0:\mu_1\leq \mu_2$ vs. $H_1: \mu_1>\mu_2$. In this setting, it is well known that under $H_0$,
\begin{equation}
\label{AncovaT}
T:= \frac{\hat{\mu}_1-\hat{\mu}_2}{\hat{\sigma}}
\end{equation}
follows the central $t$ distribution with $n_1+n_2-2-c$ degrees of freedom, where $\hat{\mu}_i$ denotes the estimated adjusted mean in group $i$, $i \in \{1,2\}$, and 
\begin{equation}
\hat{\sigma}:= \widehat{Var}(\hat{\mu}_1-\hat{\mu}_2)  
= \left(\frac{1}{n_1}+\frac{1}{n_2} + Q(Z)\right)\frac{n_1+n_2-2}{n_1+n_2-2-c} \hat{\sigma}_Y^2(1-\hat{R}^2).
\end{equation}
Thereby, $Q(Z):=\VecFormat{\bar{Z}}_{d}^{\prime}\left((n_1+n_2-2)\MatrFormatGreek{\hat{\Sigma}_{Z}}\right)^{-1}\VecFormat{\bar{Z}}_{d}$, where $\VecFormat{\bar{Z}}_d:=(\bar{Z_1}^{(1)}-\bar{Z_2}^{(1)}$, $\dots$, $\bar{Z_1}^{(c)}-\bar{Z_2}^{(c)})^{\prime}$. Moreover, $\hat{R}^2$ denotes the square of the estimated pooled multiple correlation coefficient between the outcome and the covariates, that is, $\hat{R}^2:=\VecFormatGreek{\hat{\sigma}}_{YZ}^{\prime}\MatrFormatGreek{\hat{\Sigma}}_Z^{-1}\VecFormatGreek{\hat{\sigma}}_{YZ} / \hat{\sigma}_Y^2$, where $\hat{\sigma}_Y^2$, $\VecFormatGreek{\hat{\sigma}}_{YZ}$ and $\MatrFormatGreek{\hat{\Sigma}}_Z$ denote the pooled estimators of $\sigma_Y^2$, $\VecFormatGreek{\sigma}_{YZ}$ and $\MatrFormatGreek{\Sigma}_Z$,
 respectively. Exact unconditional power for the ANCOVA $F$ test was derived by Shieh.\cite{Shi17} Following this approach, exact sample sizes for a specified power value can be obtained iteratively. However, in order to facilitate sample size calculations by avoiding numerical or iterative computations, several approximations have been proposed. In the sequel, $N = n_1 + n_2$ denotes the total sample size, $\gamma = n_2 / n_1$ the allocation ratio, $\alpha$ and $\beta$ are the type I and II error levels, and $z_p$ denotes the $p$-quantile of the standard normal distribution. Let $\Delta$ be the stipulated difference of the adjusted means (i.e., the clinically relevant difference). Then, by generalizing the approximate sample size calculation methods discussed by Friede and Kieser\citep{Kie11}, the following formulas are proposed.
 
\begin{enumerate}
\item Basic approximate formula: 
\begin{equation}
\label{Appr}
N_{A} = \frac{(\gamma+1)^2}{\gamma}\frac{(z_{1-\alpha/2}+z_{1-\beta})^2 \sigma_Y^2 (1-R^2)}{\Delta^2},
\end{equation}
\item Guenther-Schouten-like adjustment (GS): 
\begin{equation}
\label{GS}
N_{GS} = N_A + \frac{z_{1-\alpha/2}^2}{2}.
\end{equation}
\item Degrees-of-freedom adjustment (DF):
\begin{equation}
\label{DF}
N_{DF} = N_A \frac{N_A-2}{N_A-2-c}.
\end{equation}
\item Combined Guenther-Schouten and degrees-of-freedom adjustment (GS + DF):
\begin{equation}
\label{GSDF}
N_{GS,DF} = N_{DF} + \frac{z_{1-\alpha/2}^2}{2}.
\end{equation}
\end{enumerate}   

It should be noted that the sample size formulas correspond to a one-sided level $\alpha/2$ test, as recommended by guidelines regarding hypothesis testing in a superiority setting.\citep{IchE9} A formal justification of \refmath{Appr} as well as some heuristic arguments concerning \refmath{DF} and \refmath{GSDF} are provided in the appendix.

\subsection{A method for blinded interim sample size reassessment by re-estimating nuisance parameters}
\label{SubsectionBlinded}

We propose the following procedure for blinded sample size recalculation:
\begin{enumerate}
\item Calculate the initial sample size $N_{init}$, using the degrees-of-freedom adjustment approach \refmath{DF}. \label{step1}
\item As soon as data from $N_{\tau}:=\tau N_{init}$ patients is available, $\tau > 0$, estimate the residual variance $\hat{\sigma}_{\tau}^2$ of the linear regression model 
\[
Y_{ij} = \beta_0 + \sum_{k=1}^{c}Z_{ij}^{(k)}\beta_k + \epsilon_{ij}, i \in \{1,2\}, j \in \{1,2,\ldots,n_i\},
\]
that is, the residual variance based on a regression model for the pooled sample. This approach has been proposed for the case of one random covariate by Friede and Kieser \cite{Kie11}. The main idea is that unblinding is not required here, because the residual variance based on the data from the pooled sample can be calculated without knowing the group indicators. Of course, the estimators will be biased, but evidence from the simple ANCOVA suggests that the impact of that bias on the expected sample size is small.\citep{Kie13}\label{step2}
\item Recalculate the sample size by using the Guenther-Schouten approach, that is, 
\[
\hat{N}_{rec} = \frac{(\gamma+1)^2}{\gamma}\frac{(z_{1-\alpha/2}+z_{1-\beta})^2 \hat{\sigma}_{\tau}}{\Delta^2} + \frac{z_{1-\alpha/2}^2}{2}.
\]
At this point, one may ask why the Guenther-Schouten adjustment is used instead of the degrees-of-freedom adjustment. The reason is that the residual variance of a regression model with outcome $Y$ and explanatory variables $Z_1,\ldots, Z_c$ is equal to $\sigma_Y^2(1-R^2)\kappa$, where $\kappa = (N-2)/(N-2-c)$. Hence, the residual variance estimator from the previous step already accounts for the degrees of freedom. Consequently, the degrees-of-freedom adjustment lacks  any underlying rationale in this context and could even lead to an over-adjustment.   \label{step3}
\item The final sample size is determined according to the formula 
\[
\hat{N}_{final} = \min\{\max(N_{\tau},\hat{N}_{rec}),N_{bound}\}, 
\]
where $N_{bound} = k\cdot N_{init}$, $k \geq 1$, and $k$ is determined by the available resources and the time horizon for the trial. As a consequence, the final sample size is contained in the interval $[N_{\tau},k\cdot N_{init}]$. Thus, we allow for either in- or decreasing the initially planned sample size at the interim recalculation, yet making sure that the final sample size cannot grow ``too large'' compared to $N_{init}$. The latter restriction is sensible from a statistical point of view, especially when $N_{\tau}$ is quite small, because substantial uncertainty in estimating the residual variance might lead to an excessive inflation of the final sample size. Moreover, from a practical point of view, financial and human resources are limited. Actually, bounding the final sample size, for example, at the fourfold of $N_{init}$ is a quite ambitious choice, and even a doubled sample size might already present a challenge in implementation.\label{step4}\\
\end{enumerate}

Needless to say that the sample sizes obtained in each step are rounded to the smallest integer that is equal to or greater than the calculated value. In addition, depending on the allocation ratio  $\gamma$, another adjustment has to be done. For example, if we consider a balanced design (i.e., $\gamma = 1$) and the total final sample size $\hat{N}_{final}$ is odd, we must add $1$ in order to get an even number.

\subsection{Some practical considerations}
In the fixed sample size setting as well as in the first step of our proposed sample size reassessment procedure, several quantities have to be provided by subject-matter experts or have to be extracted from previous studies. When calculating approximate sample sizes for a $t$ test without any covariates, only $\sigma_Y^2$ and $\Delta$ are required. Now, the approximate formula \refmath{Appr} in the multiple ANCOVA setting differs from the $t$ test formula by the factor $(1-R^2)$. Especially for $c=1$, this is a well-known rule-of-thumb: At first, one calculates the sample size as if a $t$ test was used in the final analysis. In a second step, the sample size is being multiplied with one minus the squared correlation between the outcome and the covariate. Formula \refmath{Appr} shows that this calculation rule can be extended to the multiple ANCOVA case, with the squared correlation replaced by the squared multiple correlation coefficient $R^2$. However, in practice, it might be difficult to obtain this factor directly. Therefore, we would like to briefly discuss the following approaches:

\begin{enumerate}
\item \textit{Iterative calculation}: Apply the following formula to calculate $R^2$ iteratively, based on an $R^2$ value from a reduced model and a partial correlation coefficient 
\citep{Rav}: 
\begin{equation*}
R_{Y;Z_1,\dots,Z_c}^2 = R_{Y;Z_1,\dots,Z_{c-1}}^2 + (1-R_{Y;Z_1,\dots,Z_{c-1}}^2)\rho_{(YZ_c)|(Z_1,\dots,Z_{c-1})}^2,
\end{equation*}   
where $R^2$ and $\rho^2$ denote the squared multiple and partial correlation coefficients, respectively. For example, in case of $c=2$, this formula reduces to 
\[
R_{Y;Z_1,Z_2}^2 = \rho_{YZ_1}^2 + (1-\rho_{YZ_1}^2)\rho_{(YZ_2)|Z_1}^2.
\]
Some caution is required here: Formally, the partial correlation coefficient $\rho_{(YZ_2)|Z_1}$ is defined as the Pearson correlation based on the elements of the covariance matrix of the conditional distribution of $(Y,Z_2)$, given $Z_1 = z_1$ for some $z_1 \in \mathbb{R}$. However, that covariance matrix does not depend on the particular choice of $z_1$.\citep{Rav} Hence, one could basically use any estimated correlation between $Y$ and $Z_2$ from a previous study as a ``good guess'' of that partial correlation coefficient. \\

\item \textit{Definition of} $R^2$: Taking into account that $R^2 =  (\VecFormatGreek{\sigma}_{YZ}^{\prime}\MatrFormatGreek{\Sigma}_{Z}\VecFormatGreek{\sigma}_{YZ} / \sigma_Y^2)$, $R^2$ is calculated by specifying the $(c+1)(c+2)/2$ nuisance parameters which uniquely define the covariance matrix of the covariates as well as the correlations between the covariates and the outcome and the variance of the outcome. This approach might well be feasible in practice, because it is moreorless straightforward to extract these correlations and variances from previous studies or infer them from expert opinion.  However, apart from the considerable increase in the number of parameters for growing $c$, some configurations of individual parameters might yield an $R^2$ value that exceeds $1$, which would in turn lead to negative sample sizes. For example, consider a scenario with $c=2$ covariates, and let $\sigma_Y^2 = \sigma_{Z_1}^2 = \sigma_{Z_2}^2 = 1$, $Cov(Y,Z_1) = Cov(Y,Z_2) = 0.7$ and $\rho_{Z_1,Z_2} = Cov(Z_1,Z_2) = -0.3$. Then, $R^2 = 1.4 > 1$. Interestingly, this cannot be explained by merely referring to the fact that there might be some logical inconsistencies (observe that $Cov(Y,Z_1)$ and $Cov(Y,Z_2)$ are both positive, whereas $Cov(Z_1,Z_2)$ is negative). If we slightly change the setting and let $Cov(Y,Z_1) = Cov(Y,Z_2) = 0.5$, we get $R^2 = 0.714 < 1$. Therefore, everything should work well in this case, although there still might be some logical inconsistencies due to the different signs of the covariances. This surprising behaviour has not been noted in the literature so far, although that sort of difficulties is likely to arise in practice.\\
In order to solve this problem, one should check whether the joint covariance matrix $\MatrFormatGreek{\Sigma}$ is positive semidefinite, because this would imply $R^2\in [0,1]$. For the special case of $\MatrFormatGreek{\Sigma}$ having a compound symmetry structure (i.e., $\sigma^2:=\sigma_Y^2 = \sigma_{Z_1}^2 = \ldots = \sigma_{Z_c}^2$, $\rho:=\rho_{YZ_1} = \ldots = \rho_{YZ_c} = \rho_{Z_iZ_j}$ for all $i \neq j$), we can even characterize the positive semidefinite matrices $\MatrFormatGreek{\Sigma}$: The eigenvalues of $\MatrFormatGreek{\Sigma}$ are $\lambda_1 = \sigma^2(1+c\rho)$ with multiplicity $1$ and $\lambda_2 = \sigma^2(1-\rho)$ with multiplicity $c$. Thus, $\MatrFormat{\Sigma}$ is positive semidefinite if and only if $\rho \geq -1/c$. In more general cases, one has to do an eigenanalysis after having specified the nuisance parameters. In order to facilitate the use of the proposed sample size formulas for applied researchers, the \texttt{R} code provided in the online supplement automatically checks for positive semidefiniteness and returns a warning message if this condition is not met.

\end{enumerate}

\section{Simulation study}
\label{Simu}
In order to evaluate the performance of the aforementioned approximate fixed sample size formulas and our proposed sample size recalculation procedure, we conducted an extensive simulation study. At first, we investigated the case of $c=2$ covariates. The observations and covariates were drawn from a trivariate normal distribution with mean vectors $(\mu_i,0,0)^{\prime}$, $i \in \{1,2\}$, where $\mu_2 = 0$ and $\mu_1 = \Delta \in  \{0.25,0.5,0.75\}$. Further, we assumed a compound symmetry structure for the covariance matrix of the covariates $\MatrFormatGreek{\Sigma}_Z$, that is $\MatrFormatGreek{\Sigma}_Z = \sigma_Z^2 I_2 + \rho_Z(J_2-I_2),$
where $I_2$ and $J_2$ denote the $2$-dimensional identity matrix and the matrix containing all 1's, respectively. We set $\sigma_Y^2 = \sigma_Z^2 = 1$ and considered all combinations of $\VecFormatGreek{\sigma}_{Y\VecFormat{Z}} \in \{(0.25,0.25)',(0.5,0.5)',(0.75,0.75)', (0.25,0.5)',(0.25,0.75)',(0.5,0.75)'\}$ and $\rho_Z\in \{0.25,0.5,0.75\}$. Observe that we thus cover a broad range of $R^2$ values, since these configurations yield $R^2\in \{0.071$, $0.083$, $0.100$,
$0.250$, $0.267$, $0.286$, $0.333$, $0.400$, $0.567$, $0.571$, $0.583$, $0.643$, $0.667$, $0.750$, $0.786$, $0.900\}$. Regarding the group sizes, we considered a balanced design (i.e., $\gamma = n_2 / n_1 = 1$). Additionally, we repeated the simulations for all scenarios where $\rho_Z = 0.5$ employing a 1:2 allocation ratio (i.e., $\gamma = 2$). To our knowledge, only 1:1 allocation has been considered in the literature so far, despite the practical relevance of unequal allocation ratios. For example, in a study where the investigators want to show superiority of a new drug over the standard treatment, but the new drug is supposed to carry a high risk of severe adverse events, unequal allocation may be an attractive option. Moreover, in order to investigate whether the performance of our proposed method changes substantially with increasing $c$, we conducted simulations for $c=3$ covariates, too. We set $\sigma_{Z_1}^2 = \sigma_{Z_2}^2 = \sigma_{Z_3}^2 = 1$, $Cov(Z_1,Z_2) = Cov(Z_2,Z_3) = 0.5$ and $Cov(Z_1,Z_3) = 0.25$. All other specifications were the same as described above, with $Cov(Y,Z_3) = 0.5$ as the third coordinate of $\VecFormatGreek{\sigma}_{YZ}$.
For each scenario, we conducted $n_{sim} = 1,000,000$ simulation runs, which resulted in an estimated standard error of $0.0004$ for the empirical power ($1-\beta = 0.80$) and $0.0002$ for the type I error rate ($\alpha/2 = 0.025$), respectively. \\

At first, we report the results concerning the performance of the four fixed sample size procedures in terms of sample sizes and power, in comparison with the results from the exact approach proposed by Shieh.\cite{Shi17} The basic approximate formula \refmath{Appr} performed well only in large samples, yet showing considerable deviations from the respective exact values in small samples. Proceeding with formulas \refmath{GS}--\refmath{GSDF}, the performance was gradually getting better, with the sample sizes based on the combined Guenther-Schouten and degrees-of-freedom adjustment \refmath{GSDF} being equal to the exact sample sizes in most scenarios. As an alternative, however, the degrees-of-freedom adjustment \refmath{DF} also yielded good approximations, unless the sample sizes were very small. Apart from that, it should be noted that the sample sizes were larger for unbalanced scenarios (i.e., $\gamma = 2$) than in balanced settings, regardless whether the exact or one of the approximate methods was used.  All results can be found in full detail in Tables 1--5 in the online supplement. \\   

\begin{figure}[ht]
\centering
\includegraphics[width=14cm, keepaspectratio]{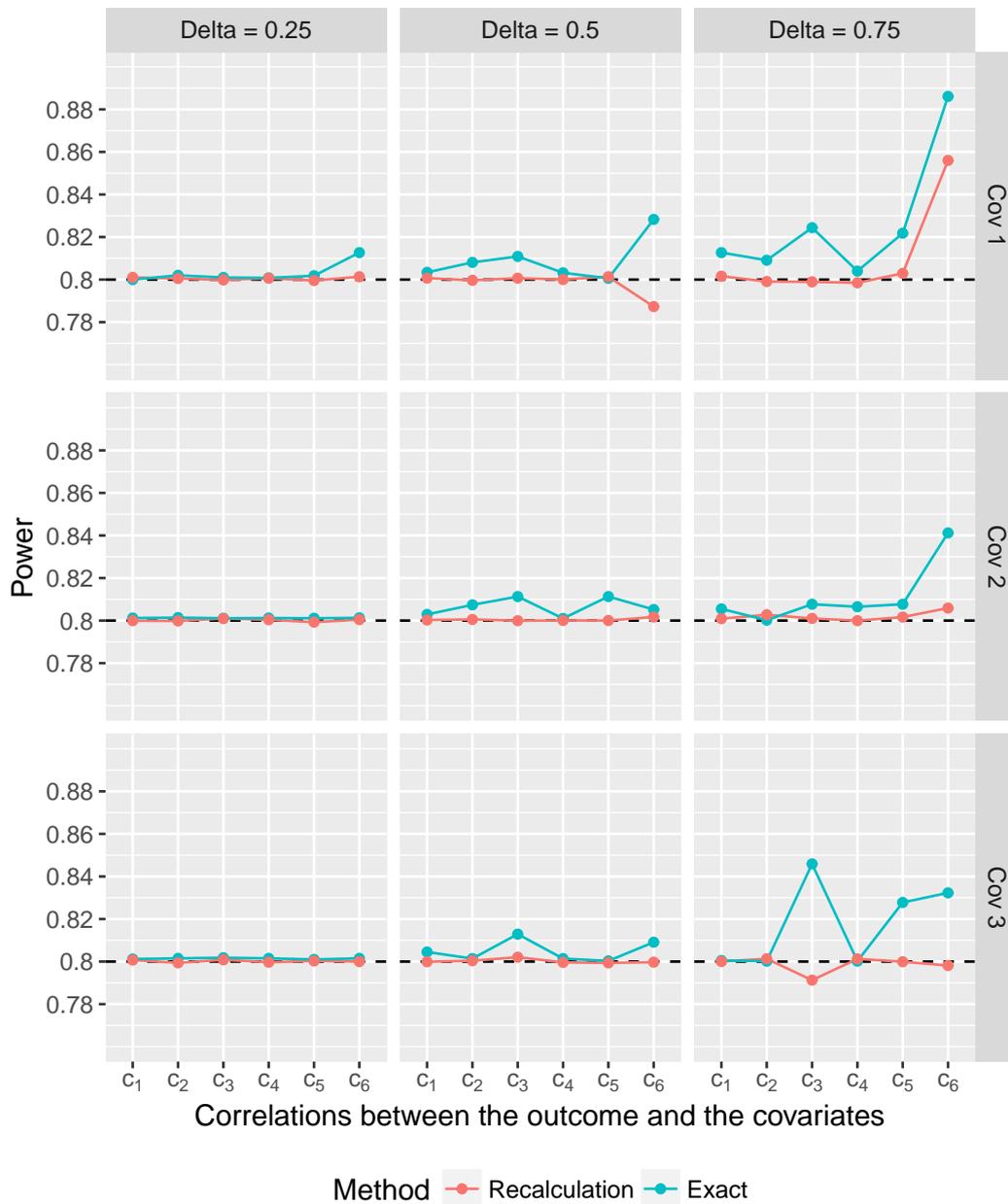}
\caption{Empirical power of the proposed sample size recalculation procedure vs. exact power in the fixed sample size design with correct specification of nuisance parameters, and target power. Cov 1 to 3 indicate different compound symmetry structures of the covariance matrix of the covariates ($\rho_Z = i\times 0.25, i \in \{1,2,3\}$), and the labels on the x axis denote the correlations between the outcome and covariate $j$, $j \in \{1,2\}$: $c_1 = (0.25,0.25)$, $c_2 = (0.5,0.5)$, $c_3 = (0.75,0.75)$, $c_4 = (0.25,0.5)$, $c_5 = (0.25,0.75)$, $c_6 = (0.5,0.75)$. \label{Fig1}}
\end{figure}

For the simulations regarding our proposed sample size recalculation procedure, the parameter $\tau$, which specifies the time point of the interim reassessment, was set to $0.5$. The final total sample size was bounded at the fourfold of the initially planned sample size $N_{init}$ (i.e., $k = 4$). Final average sample sizes and empirical power were compared to the corresponding values that were obtained by using the exact fixed sample size calculation approach proposed by Shieh.\cite{Shi17} Firstly, in balanced designs with $c=2$ covariates, the pre-specified level of $\alpha/2 = 0.025$ was well maintained, with a median type I error rate of $0.02509$ (range $0.02462 - 0.02554$). 
The median simulated power of the recalculation procedure was $0.80028$ (range $0.78731 - 0.85603$). The exact sample sizes corresponding to the minimum and maximum simulated power values are $N = 18$ and $N = 12$, respectively, indicating that the method might have a somewhat suboptimal performance for very small total sample sizes. This corresponds to a trend towards increased deviations from the target power for large hypothesized effects $\Delta$ (Figure \ref{Fig1}). At this point, however, the question arises if it is sensible to consider a scenario with extremely small sample sizes at all, because in the interim reassessment, a regression model with $3$ parameters is fitted to data from $N/2\in \{6,9\}$ subjects.

Anyway, apart from these few instances, the proposed sample size recalculation procedure performs very well, with the empirical power close to the target level. For example, if only those scenarios with a total exact sample size $\geq 30$ (i.e., $15$ subjects per group) are considered, the empirical power of the recalculation procedure ranges from $0.79850$ to $0.80272$. Moreover, observe that the simulated power for the recalculation procedure is most of the time closer to the target level of $0.8$ than the exact power value, even though all parameters were correctly specified. With respect to the sample sizes, the recalculation procedure yields expected final sample sizes which exceed the exact values by $6.1$ subjects on average. Depending on the particular setting, the difference can be up to $7.1$. Similar discrepancies have been found for the ANCOVA model with one covariate, as reported by Friede and Kieser.\cite{Kie11} So, even in the case of $c=2$ covariates, the overall ``price'' one has to pay for the increased flexibility of the recalculation procedure is small, thus rendering the proposed method useful for practical applications. However, one has to keep in mind that especially in small samples, the maximum final sample size resulting from the recalculation procedure could exceed the initially planned sample size considerably. 

\begin{figure}[ht]
\centering
\includegraphics[width=14cm, keepaspectratio]{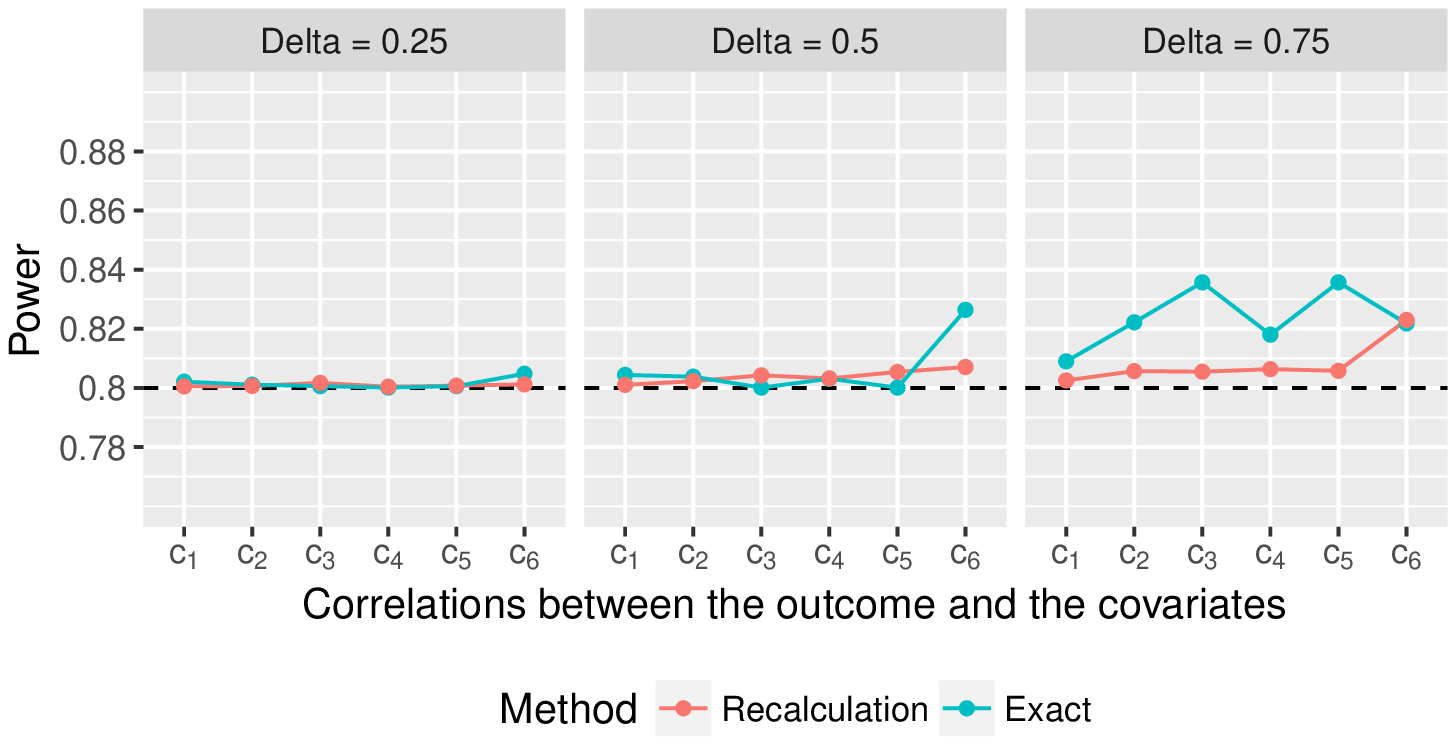}
\caption{Empirical power of the proposed sample size recalculation procedure vs. exact power in the fixed sample size design with correct specification of nuisance parameters, and target power, assuming a compound symmetry structure of the covariance matrix of the covariates, where $\sigma_Z = 1$ and $\rho_Z = 0.5$, and unbalanced group sizes, $n_2 = 2 n_1$. The labels on the x axis denote the correlations between the outcome and covariate $j$, $j \in \{1,2\}$: $c_1 = (0.25,0.25)$, $c_2 = (0.5,0.5)$, $c_3 = (0.75,0.75)$, $c_4 = (0.25,0.5)$, $c_5 = (0.25,0.75)$, $c_6 = (0.5,0.75)$ \label{Fig2}}
\end{figure}

Basically, compared to the balanced settings, the performance for $\gamma = 2$ is similar, if not even slightly better. The empirical type I error rates are close to the nominal level (median $0.02506$, range $0.02456 - 0.02558$). The maximum deviations from the target are even smaller than in the balanced settings, and the empirical power always lies above $0.8$ now (median empirical power $0.80287$, range $0.80041 - 0.82300$). Especially for $\Delta = 0.75$, the exact power substantially exceeds the target level, whereas the deviation of the power of the recalculation method is much lower (Figure \ref{Fig2}). The average difference between the expected final sample sizes and the corresponding fixed exact sample sizes is $6.0$ (range $4.4 - 7.4$), which is similar to the balanced setting, too. \\

\begin{figure}[ht]
\centering
\includegraphics[width=14cm, keepaspectratio]{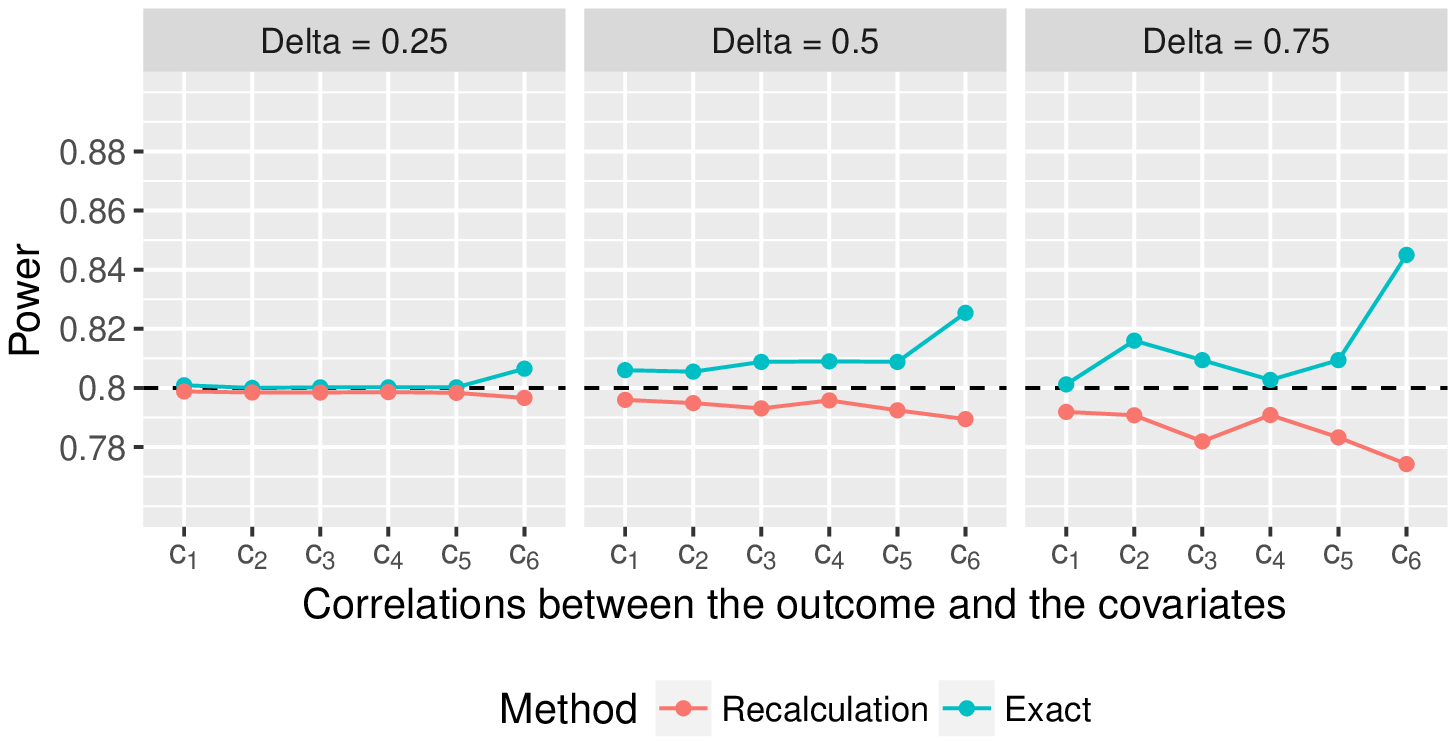}
\caption{Empirical power of the proposed sample size recalculation procedure vs. exact power in the fixed sample size design with correct specification of nuisance parameters, and target power, assuming an ANCOVA model with $c = 3$ covariates, a compound symmetry structure of the covariance matrix of the covariates, where $\sigma_Z = 1$ and $\rho_Z = 0.5$, and balanced group sizes, $n_1 = n_2$. The labels on the x axis denote the correlations between the outcome and covariate $j$, $j \in \{1,2,3\}$: $c_1 = (0.25,0.25,0.5)$, $c_2 = (0.5,0.5,0.5)$, $c_3 = (0.75,0.75,0.5)$, $c_4 = (0.25,0.5,0.5)$, $c_5 = (0.25,0.75,0.5)$, $c_6 = (0.5,0.75,0.5)$ \label{Fig3}}
\end{figure}

In case of balanced designs with $c=3$ covariates, the proposed sample size reassessment procedure yields empirical power values which are consistently smaller than $0.8$. While the method still performs well for $\Delta \in \{0.25, 0.5\}$, there is some power loss for $\Delta = 0.75$ in most cases (e.g., empirical power of $0.77424$ for $\VecFormatGreek{\sigma}_{YZ} = (0.75,0.75,0.5)'$; Figure \ref{Fig3}). However, the type I error rates are still close to the pre-specified level (median $0.02501$, range $0.02472 - 0.02527$).\\

\begin{figure}[ht]
\centering
\includegraphics[width=14cm, keepaspectratio]{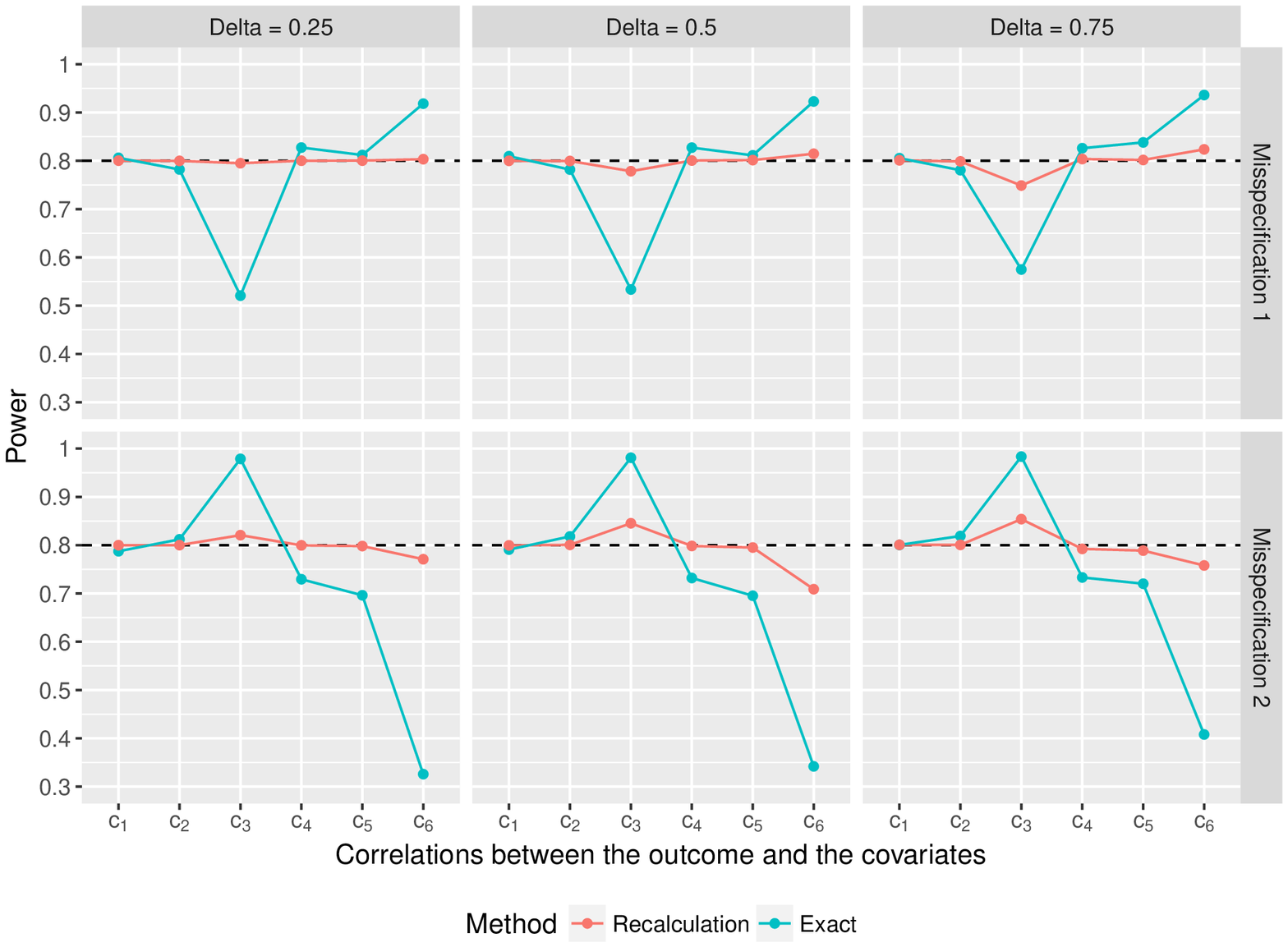}
\caption{Empirical power of the proposed sample size recalculation procedure vs. exact power in the fixed sample size design in case of $c=2$ covariates, assuming $\sigma_Z = 1$, and balanced group sizes, $n_1 = n_2$. The labels on the x axis denote the correlations between the outcome and covariate $j$, $j \in \{1,2\}$: $c_1 = (0.25,0.25)$, $c_2 = (0.5,0.5)$, $c_3 = (0.75,0.75)$, $c_4 = (0.25,0.5)$, $c_5 = (0.25,0.75)$, $c_6 = (0.5,0.75)$. Misspecification 1: $\rho_Z = 0.75$ was assumed for sample size planning, although the true correlation was $0.5$. Misspecification 2: $\rho_Z = 0.25$ was assumed for sample size planning, although the true correlation was $0.75$. \label{Fig4}}
\end{figure}

So far, we have only considered somewhat idealistic settings, assuming that the nuisance parameters were correctly specified in advance. In these scenarios, where a fixed sample size calculation approach is supposed to yield reliable results, the recalculation procedure performed equally well and was even slightly superior in some cases. Additionally, we shall consider two settings now, where the advantages of the increased flexibility due to the interim reassessment becomes even more obvious, because the covariance matrix $\Sigma_Z$ of the covariates was misspecified. On the one hand, we assumed that initially, $\rho_Z$ was set to $0.75$, but the true value was $0.5$. On the other hand, instead of $\rho_Z = 0.75$, we used $\rho_Z = 0.25$ for the initial sample size calculation. 
It can be seen in Figure \ref{Fig4} that our proposed recalculation procedure outperformed the fixed sample size calculation method in all scenarios. The latter deviated substantially from the target power of 80 percent in some settings. Although the sample size reassessment procedure was somewhat underpowered for covariance scenario $c_6$ (i.e., moderate to strong correlations between the outcomes and the covariates, which in turn translates to small sample sizes) and misspecification scenario 2 (i.e., assuming that $\rho_Z$ was equal to $0.25$ instead of $0.75$), it is still preferable over the fixed sample size calculation method, which totally failed in that scenario. Moreover, similarly to all previously discussed settings, our proposed method again keeps the $\alpha$ level well (median 0.02498, range 0.02466--0.02568).
  
\section{Application to a real-life setting}
\label{RealLife}
In order to illustrate the application of our proposed sample size recalculation procedure, we use data from the SIESTA (Sedation vs. Intubation for Endovascular Stroke TreAtment) trial. In this monocentric randomized parallel-group trial, the main aim was to assess if conscious sedation is superior to general anesthesia for early neurological improvement among patients receiving stroke thrombectomy. The change from baseline of the National Institutes of Health Stroke Scale (NIHSS) 24 hours after the intervention was considered as the primary outcome, which was compared between the two groups while adjusting for the NIHSS at baseline. For further details, we shall refer to the original publication of the results.\citep{Sch16} Regarding sample size calculation, the investigators report that the estimated variances, which were obtained from previous studies, showed considerable variation. Even worse, an estimate of the correlation between the outcome and the NIHSS at baseline was not available at all. Therefore, it was decided to conduct a preliminary sample size calculation with the t test formula, which yielded a total sample size $N = 100$, assuming equal group allocation. When data of $75$ patients was available (i.e., $\tau = 0.75$), a pre-planned blinded sample size recalculation was carried out, using the method proposed by Friede $\&$ Kieser.\citep{Kie11} Further details are provided in the study protocol.\citep{Sch15}\\
Clearly, this is a case where an interim sample size reassessment is more appropriate than calculating sample sizes in advance and keeping them fixed throughout the conduct of the study. The latter approach would have to rely on a somewhat arbitrary initial guess of the correlation and on a variance estimate which is subject to considerable uncertainty. So, the method employed in the SIESTA study was definitely a good choice. Nevertheless, we shall examine whether the final sample size would have substantially changed if multiple covariates had been considered. At this point, we would like to mention that the primary outcome as well as some of the covariates, which will be considered below, are ordinally scaled variables. However, for the classical ANCOVA as well as for our proposed method to be applied, normally distributed variables are required. Since there is some evidence that type I error rates and power are not substantially affected by applying the classical ANCOVA to ordinally scaled outcomes and baseline measurements thereof \citep{Sul03}, it is appropriate to use the data for illustrating the application of our proposed sample size recalculation method. Nevertheless, we would like to emphasize that the resulting effect sizes (i.e., adjusted mean differences) should be interpreted with caution if the outcome and / or the covariates are ordinally scaled. \\
In the sequel, we perform several sample size recalculations, based on the data that was available at the time of the interim analysis, assuming a group allocation ratio $\gamma = 1$ and a clinically relevant difference $\Delta = 4$. We set $\alpha = 0.05$ and $1-\beta = 0.9$. These specifications correspond to the setup described in the study protocol; by contrast, however, we do not impose an upper bound on the final sample size, because this would complicate the assessment of potential differences between the respective models. For sample size recalculation, we conducted steps 2 and 3 from Section \ref{SubsectionBlinded}. At first, we reproduced the sample size reassessment that had been actually carried out in the study, assuming an ANCOVA model with one single covariate, namely NIHSS at baseline. The estimated residual variance and final sample size were $\hat{\sigma}_{\tau} = 99.35$ and $\hat{N}_{rec} = 264$, respectively. Next, we added the age of the patients (in years) as a second covariate. Using this model, both values slightly decreased ($\hat{\sigma}_{\tau} = 96.99$, $\hat{N}_{rec} = 258$). As an alternative, we considered replacing age by the degree of recanalization, as quantified by the Thrombolysis in cerebral infarction (TICI) scale. So, again, we had a scenario with $c=2$ covariates, namely NIHSS at baseline and TICI. Obviously, the inclusion of TICI led to a substantial decrease of the final sample size down to $\hat{N}_{rec} = 214$ ($\hat{\sigma}_{\tau} = 80.42$). Finally, we added the door-to-intervention time (in seconds) to the model. Taking NIHSS at baseline, TICI and door-to-intervention time as covariates, we got $\hat{\sigma}_{\tau} = 77.43$ and $\hat{N}_{rec} = 206$, respectively. \\
Summing up, we would like to emphasize three important points. Firstly, we have seen that in all scenarios, the recalculated sample sizes substantially exceeded the initially planned sample size $N=100$. So, obviously, if the reassessment had not been applied, the study would have been considerably underpowered. Secondly, careful thoughts about the associations between the outcome and potential covariates are required in the planning phase. In the example, the inclusion of TICI led to a considerable decrease of the final sample size, whereas there was hardly any use in adding age to the model. Thirdly, although our example of course only provides limited evidence, it seems as if increasing the number of covariates beyond $c=2$ would not lead to further substantial improvements, which is consistent with findings in a similar context (repeated measures).\citep{Fri92}

\section{Discussion and conclusions}
\label{Discussion}

Although some work concerning exact and approximate sample size calculation in the context of an ANCOVA model with multiple covariates has been published recently, there is currently no method available, which allows for blinded sample size recalculation in that setting. We have proposed such a method here, based on the approach suggested by Friede and Kieser \cite{Kie11} for the situation of one single covariate. At the time point of the blinded interim reassessment, the residual variance is estimated based on the pooled interim data, and the final sample size is recalculated accordingly. The results of our extensive simulation study indicate that even if all parameters are correctly specified, the performance of the blinded recalculation method is close to the exact results for models with 2 random covariates and for most scenarios with 3 covariates, except in cases where the required sample size is very small (i.e., $\Delta = 0.75$). In more realistic scenarios, where initial misspecifications of a nuisance parameter are present, our proposed method is clearly superior to the fixed approach. Apart from that, regardless whether a fixed sample size calculation formula or a recalculation method is employed, caution is needed when initially specifying the values of the nuisance parameters, in order to prevent the sample sizes provided by the formula from being negative. This surprising discovery is important and has to be taken into consideration when applying the aforementioned methods in real-life settings (e.g., by checking for positive semidefiniteness of the joint covariance matrix). \\

The sample size calculation formulas, which are used in our proposed procedure, have been derived assuming a multivariate normal distribution of the covariates and the outcome. However, in applied research, this assumption is frequently violated. The outcome and / or the covariates may not even be continuous (e.g., ordinal scores like the modified Rankin Scale, the Hamilton Depression Score, etc.). We have already touched this issue briefly in Section \ref{RealLife}, providing some explanation how the ANCOVA results can be used and interpreted, though. In such instances, however, using nonparametric ANCOVA methods \citep{Bat03} would be an attractive alternative with respect to the interpretation and the robustness of the results. Nevertheless, for parallel-group comparisons, sample size calculation methods are only available for the unadjusted Wilcoxon-Mann-Whitney test  \citep{Noe87, Gov07}. Therefore, investigating sample size (re-)calculation procedures for nonparametric ANCOVA is a promising goal of future research. \\

Basically, our proposed recalculation method can be applied to ANCOVA models with an arbitrary number of covariates. However, we restricted to a thorough examination of various settings with $c=2$ and $c=3$ as well as balanced and unbalanced scenarios. Evidence from repeated measures models indicates that if the number of baseline visits is increased, there is always a reduction of the required sample size (i.e., a gain in power). However, the magnitude of that reduction decreases considerably with a growing number of baseline visits.\citep{Fri92} In our simulation study, we also noticed that already for $c=3$ the reduction in sample sizes compared to a similar scenario for $c=2$ was not larger than 10 in most cases. Moreover, especially in small to moderate samples, multicollinearity issues are more likely to occur with an increasing number of covariates. 
Therefore, all in all, we recommend including at most $3$ covariates into the model, thus maintaining a balance between gains in power and feasibility in practical applications.\\

It should generally be noted that any blinded reassessment procedure cannot correct for misspecified effect sizes. For this purpose, an adaptive design with unblinded interim analysis would be an appealing alternative. However, such a method has the disadvantage that unblinding and, thus, establishing an independent data monitoring committee are required. Furthermore, regulatory guidelines prefer blinded sample size reassessment due to avoiding bias.\citep{Ema07, Fda16}\\

To conclude, firstly, we have assessed the performance of several approximate fixed sample size calculation approaches in terms of sample sizes and power. According to the results of the simulation studies, the proposed adjustments which take the number of covariates into account are easy to apply and can be safely used in practice. Moreover, most importantly, we have proposed a blinded sample size recalculation method for an ANCOVA model with multiple random covariates and showed by extensive simulations that it maintains the pre-specified type I error level and power very well in models with $2$ or $3$ covariates, except for very small sample sizes. In case of initial misspecifications, our proposed method outperforms the fixed sample size calculation approach. Thus, applied researchers now have a procedure at hand, which is easy to use and increases the robustness of sample size calculation while at the same time keeping the treatment allocation blinded.

\section*{Appendix}

\textit{Derivation of the basic approximate sample size formula \refmath{Appr} in Section \ref{FormulasRealLife}}\\

\noindent
At first, we rewrite the test statistic $T$ defined in \refmath{AncovaT} as 
\begin{equation*}
T = \frac{\hat{\Delta}}{\sqrt{(n_1^{-1}+n_2^{-1}+\hat{Q}_Z)\kappa \hat{Q}_{YZ}}},
\end{equation*}

where $\hat{Q}_{Z}:= \VecFormat{\bar{Z}}_{d}^{\prime}\left((n_1+n_2-2)\MatrFormatGreek{\hat{\Sigma}}_{Z}\right)^{-1}\VecFormat{\bar{Z}}_{d}$, $\hat{\Delta}:=\hat{\mu}_{1}-\hat{\mu}_{2}$, $\kappa:=(n_1+n_2-2)/(n_1+n_2-2-c)$, $\hat{Q}_{YZ}:= \hat{\sigma}_Y^2(1-\hat{\sigma}_Y^{-2}\VecFormatGreek{\hat{\sigma}}_{YZ}^{\prime}\hat{\Sigma}_Z^{-1}\VecFormatGreek{\hat{\sigma}}_{YZ})$. 

Next, we apply the normal approximation of the $t$ distribution. Using $n_2 = \gamma n_1$ and doing some algebra yields 
\begin{align}
\frac{\hat{\Delta}^2}{(n_1^{-1}+n_2^{-1}+\hat{Q}_Z)\kappa \hat{Q}_{YZ}} &\approx (z_{1-\alpha/2}+z_{1-\beta})^2 \nonumber\\
\Leftrightarrow \frac{\hat{\Delta}^2}{(\gamma n_1)^{-1}(\gamma + 1 + \gamma n_1 \hat{Q}_Z)} &\approx (z_{1-\alpha/2}+z_{1-\beta})^2\kappa \hat{Q}_{YZ}\nonumber\\
\Leftrightarrow \gamma n_1\hat{\Delta}^2 &\approx (z_{1-\alpha/2}+z_{1-\beta})^2\kappa \hat{Q}_{YZ}(\gamma + 1 + \gamma n_1 \hat{Q}_Z)\nonumber\\
\Leftrightarrow n_1(1-(z_{1-\alpha/2}+z_{1-\beta})^2\kappa \hat{Q}_{YZ}\hat{Q}_Z/\hat{\Delta}^2) &\approx (z_{1-\alpha/2}+z_{1-\beta})^2\kappa \hat{Q}_{YZ}(\gamma + 1)/(\gamma\hat{\Delta}^2).\label{DerivationFormula1}
\end{align}

Now, observe that $\hat{Q}_Z$ is most likely close to 0: $\VecFormat{\bar{Z}}_{d}$ is supposed to be small for reasonable sample sizes, since the covariate means of the populations are equal to $\VecFormatGreek{\mu}_Z$ for both groups. Moreover, the elements of the inverse of $\MatrFormatGreek{\hat{\Sigma}}_Z$ will be small, too, unless the variances are close to 0, or the covariates exhibit strong linear dependencies. As either scenario would lead to potentially serious problems regarding inference, these cases can be excluded. Moreover, observe that the factor $(n_1+n_2-2)^{-1}$ leads to a further deflation of the quantity $\hat{Q}_Z$. Hence, it might be appropriate to drop the term $(z_{1-\alpha/2}+z_{1-\beta})^2\kappa\hat{Q}_{YZ}\hat{Q}_{Z}/\hat{\Delta}^2$. Thus, we can further simplify \refmath{DerivationFormula1} to
\[
n_1 \approx \frac{(z_{1-\alpha/2}+z_{1-\beta})^2 \kappa \hat{Q}_{YZ}(\gamma +1)}{\gamma\hat{\Delta}^2}.
\]
Finally, observe that $\kappa \approx 1$ for reasonably large sample sizes, and that $\hat{Q}_{YZ} = \hat{\sigma}_Y^2(1-\hat{\sigma}_Y^{-2}\VecFormatGreek{\hat{\sigma}}_{YZ}^{\prime}\hat{\Sigma}_Z^{-1}\VecFormatGreek{\hat{\sigma}}_{YZ})$, according to the definition of $\hat{Q}_{YZ}$. This completes the derivation of \refmath{Appr}.\\[1cm]

\noindent
\textit{Some remarks regarding formulas \refmath{DF} and \refmath{GSDF}}\\

The approximate formulas \refmath{DF} and \refmath{GSDF} can be motivated heuristically: In the derivation of \refmath{Appr}, the term $(N - 2)/(N - 2- c)$ is dropped. Although the impact on the results is most likely negligible for moderate to large sample sizes, it could be sensible to do a corresponding post-hoc adjustment of the approximate sample size $N_A$. It should be noted, however, that in general the factor $(N_A - 2)/(N_A - 2 - c)$ is not equal to $(N-2) / (N-2-c)$, because $N_A$ is not necessarily equal to $N$. Moreover, the DF adjustment is ``stronger'' than the GS adjustment for $c\geq 2$ and typical choices of $\alpha$, because 
\begin{equation*}
N_{DF} = N_A + c + \frac{c^2+2c}{N_A-2-c}.
\end{equation*} 
For example, for $\alpha = 0.05$ we have 
\[
z_{1-\alpha/2}^2 / 2 \approx 1.921 < 2 \leq c
\]
for $c\geq 2$. It can also be seen from the previous calculations that the difference might actually be small in most practically relevant settings, especially as upward rounding has to be taken into account, too. Therefore, it could be sensible to apply an even stronger adjustment of $N_A$ by combining GS and DF, as proposed in \refmath{GSDF}. Anyway, a common feature of \refmath{DF} and \refmath{GSDF} is that, in contrast to the GS adjustment, the number of covariates is taken into account. It has been argued that one of the major drawbacks of using \refmath{Appr} or \refmath{GS} was that the number of covariates did not play a role.\citep{Shi17} Therefore, it seems plausible that the performance can be improved by applying adjustments, which take the number of covariates $c$ into account.   






\subsection*{Conflict of interest}

The authors declare no potential conflict of interests.

\section*{Supporting information}

The following supporting information is available as part of the online article:\\

\noindent
\textbf{SSRE\_multiple\_Ancova.R} This file contains the implementation of the proposed re-calculation procedure (including a check of initially specified parameters) in \texttt{R}.\\
\noindent
\textbf{SSRE\_multiple\_Ancova\_Supplement.pdf} This document contains simulation results for the fixed sample size calculation settings discussed in the manuscript.

\bibliography{IPSAncovaBibl}

\end{document}